\documentclass[prd,twocolumn,amsmath,showpacs,superscriptaddress,amssymb,floatfix,nofootinbib]{revtex4}
\usepackage{amsfonts,graphicx,multirow,amsmath,color,xcolor,ulem}
\usepackage[dvipdfm, colorlinks=true,citecolor=blue,linkcolor=blue,urlcolor=blue]{hyperref}

\allowdisplaybreaks
\begin{document}

\title{Revisiting $D$-meson twist-2, 3 distribution amplitudes}
\author{Tao Zhong\footnote{Corresponding author}}
\email{zhongtao1219@sina.com}
\address{Department of Physics, Guizhou Minzu University, Guiyang 550025, P.R. China}
\author{Dong Huang}
\address{Department of Physics, Guizhou Minzu University, Guiyang 550025, P.R. China}
\author{Hai-Bing Fu}
\email{fuhb@cqu.edu.cn}
\address{Department of Physics, Guizhou Minzu University, Guiyang 550025, P.R. China}
\address{Department of Physics, Chongqing University, Chongqing 401331, P.R. China}

\date{\today}
\begin{abstract}
Due to the significant difference between the experimental measurements and the theoretical predictions of standard model (SM) for the value of $\mathcal{R}(D)$ of the semileptonic decay $B\to D\ell\bar{\nu}_{\ell}$, people speculate that it may be the evidence of new physics beyond the SM. Usually, the $D$-meson twist-2, 3 distribution amplitudes (DAs) $\phi_{2;D}(x,\mu)$, $\phi_{3;D}^p(x,\mu)$ and $\phi_{3;D}^\sigma(x,\mu)$ are the main error sources when using perturbative QCD factorization and light-cone QCD sum rules to study $B\to D\ell\bar{\nu}_{\ell}$. Therefore, it is important to get more reasonable and accurate behaviors for those DAs. Motivated by our previous work [Phys. Rev. D 104, no.1, 016021 (2021)] on pionic leading-twist DA, we revisit $D$-meson twist-2, 3 DAs $\phi_{2;D}(x,\mu)$, $\phi_{3;D}^p(x,\mu)$ and $\phi_{3;D}^\sigma(x,\mu)$. New sum rules formulae for the $\xi$-moments of these three DAs are suggested to obtain more accurate values. The light-cone harmonic oscillator models for those DAs are improved, and whose model parameters are determined by fitting the values of $\xi$-moments with the least squares method.
\end{abstract}

\maketitle

\section{introduction}
Since 2012, semileptonic decay $B\to D\ell\bar{\nu}_\ell$ has been considered as one of the processes most likely to prove the existence of new physics beyond the standard model (SM). The reason is well known, that is, the significant difference between the experimental measurements of the ratio $\mathcal{R}(D)$ and its theoretical predictions of SM. The latest statistics given by Heavy Flavor Average Group website~\cite{HFLAV:2019otj} shows that the experimental average value of $\mathcal{R}(D)$ is $\mathcal{R}^{\rm exp.}(D) = 0.339 \pm 0.026 \pm 0.014$, while its average value of SM predictions is $\mathcal{R}^{\rm the.}(D) = 0.300 \pm 0.008$~\cite{FlavourLatticeAveragingGroup:2019iem}. The former comes from the experimental measurements for semileptonic decay $B\to D\ell\bar{\nu}_\ell$ by BaBar Collaboration in 2012~\cite{BaBar:2012obs} and 2013~\cite{BaBar:2013mob}, by Belle Collaboration in 2015~\cite{Belle:2015qfa} and 2019~\cite{Belle:2019rba}. The later is obtained by combining two lattice calculations by MILC Collaboration~\cite{MILC:2015uhg} and HPQCD Collaboration~\cite{Na:2015kha}. The authors of Ref.~\cite{Bigi:2016mdz} fit experimental and lattice results for $B\to D\ell\bar{\nu}_\ell$ to give $\mathcal{R}(D) = 0.299 \pm 0.003$. Within the framework of the Heavy-Quark Expansion, Ref.~\cite{Bordone:2019vic} gives $\mathcal{R}(D) = 0.297 \pm 0.003$. By fitting the experimental data, lattice QCD and QCD sum rules (SRs) results for $\bar{B}\to D\ell\bar{\nu}_\ell$, Ref.~\cite{Bernlochner:2017jka} predicts $\mathcal{R}(D) = 0.299 \pm 0.003$. Along with the experimental data, Ref.~\cite{Jaiswal:2017rve} use the lattice predictions~\cite{MILC:2015uhg,Na:2015kha} for the form factors of $B\to D\ell\bar{\nu}_\ell$ as inputs, the prediction for $\mathcal{R}(D)$ with the  Caprini-Lellouch-Neubert parameterization~\cite{Caprini:1997mu} of the form factors is given by $\mathcal{R}(D) = 0.302 \pm 0.003$, while using  Boyd-Grinstein-Lebed parameterization~\cite{Boyd:1997kz}, the authors obtain $\mathcal{R}(D) = 0.299 \pm 0.004$. Earlier, based on the heavy quark effective theory (HQET), Refs.~\cite{Tanaka:2010se,Fajfer:2012vx} predict $\mathcal{R}(D) = 0.302 \pm 0.015$. By using light-cone sum rules (LCSRs) approach with $B$-meson distribution amplitudes (DAs) in HQET, Ref.~\cite{Wang:2017jow} gives $\mathcal{R}(D) = 0.305^{+0.022}_{-0.025}$ in 2017. Based on the $D$-meson DAs~\cite{Zhang:2017rwz,Zhong:2018exo} obtained by QCD SRs in the framework of background field theory (BFT)~\cite{Huang:1989gv,Zhong:2014jla}, our previous work gets $\mathcal{R}(D) = 0.320^{+0.018}_{-0.021}$~\cite{Zhong:2018exo} with LCSRs.

In $B\to D$ semileptonic decay and other $D$-meson related processes, $D$-meson twist-2 DA $\phi_{2;D}(x,\mu)$, twist-3 DAs $\phi_{3;D}^p(x,\mu)$ and $\phi_{3;D}^\sigma(x,\mu)$ are usually the key input parameters and the main error sources. Among them, there are relatively more studies on leading-twist DA $\phi_{2;D}(x,\mu)$, such as, earlier, the  Kurimoto-Li-Sanda (KLS) model~\cite{Kurimoto:2002sb} based on the expansion of the Gegenbauer polynomials, the Li-L\"{u}-Zou (LLZ) model~\cite{Li:2008ts} by considering a sample harmonic-like $k_\perp$-dependence on the basis of KLS model, the Gaussian-type Li-Melic (LM) model~\cite{Li:1999kna} by employing the solution of a relativistic scalar harmonic oscillator potential for the orbital part of the wavefunction (WF), the light-cone harmonic oscillator (LCHO) model~\cite{Guo:1991eb,Zuo:2006re} based on the Brodsky-Huang-Lepage (BHL) prescription~\cite{BHL}, etc. Recently, in 2019, Ref.~\cite{Dhiman:2019ddr} studied $D$-meson twist-2 DA $\phi_{2;D}(x,\mu)$ with the light-front quark model (LFQM) by adopting the Coulomb plus exponential-type confining potential, and given the values of whose first six $\xi$-moments. Our research on $D$-meson twist-2 DA $\phi_{2;D}(x,\mu)$ is in 2018~\cite{Zhang:2017rwz}. We studied $\phi_{2;D}(x,\mu)$ by combining phenomenological LCHO model and non-perturbative QCD SRs approach. By introducing longitudinal WF $\varphi_{2;D}(x)$, we improved the LCHO model of $\phi_{2;D}(x,\mu)$ proposed in Refs.~\cite{Guo:1991eb,Zuo:2006re}. The behavior of our DA is determined by the first four Gegenbauer moments. Those Gegenbauer moments were calculated with QCD SRs in the framework of BFT. Subsequently, in the same year, we used the same method to study $D$-meson twist-3 DAs $\phi_{3;D}^p(x,\mu)$ and $\phi_{3;D}^\sigma(x,\mu)$ and further studied $B\to D$ transition form factors (TFFs) with LCSRs and calculated $\mathcal{R}(D)$~\cite{Zhong:2018exo}.

Last year we proposed a new scheme to study pionic leading-twist DA $\phi_{2;\pi}(x,\mu)$ in Ref.~\cite{Zhong:2021epq}. Firstly, we suggested a new sum rule formula for $\xi$-moment of $\phi_{2;\pi}(x,\mu)$ based on the fact that the sum rule of zeroth moment can not be normalized in entire Borel parameter region. Secondly, we adopted the least squares method to fit the values of the first ten $\xi$-moments to determine the behavior of $\phi_{2;\pi}(x,\mu)$. In fact, there are several other approaches, such as traditional QCD sum rules~\cite{Ball:2003sc}, Dyson-Schwinger equation~\cite{Chang:2013pq}, lattice calculation~\cite{RQCD:2019osh, LatticeParton:2022zqc}, etc., to be adopted in the study of the DAs of mesons especially light mesons. By comparison, the scheme suggested in Ref.~\cite{Zhong:2021epq} has its own unique advantages. In which, the new sum rule formula of $\xi$-moment can reduce the system uncertainties caused\footnote{The numerical results in Ref.~\cite{Zhong:2021epq} show that this improves the accuracy of $\xi$-moments by at least $10\%$.} by the truncation of the high-dimensional condensates as well as the simple parametrization of quark-hadron daulity for continuum states, and this improves the prediction accuracy of QCD SRs and its prediction ability for higher moments; The least squares method is used to fit the $\xi$-moments to determine DA, which avoids the extremely unreliable high-order Gegenbauer moments, and can absorb as much information of DA carried by high-order $\xi$-moments as possible to give more accurate behavior of DA~\cite{Zhong:2022lmn}. Very recently, this scheme has been used to study the kaon leading-twist DA $\phi_{2;K}(x,\mu)$ by considering the $SU_f(3)$ symmetry breaking effect~\cite{Zhong:2022ecl}, the axial-vector $a_1(1260)$-meson longitudinal twist-2 DA~\cite{Hu:2021lkl}, the scaler $K_0^\ast(1430)$ and $a_0(980)$-meson leading-twist DAs~\cite{Huang:2022xny, Wu:2022qqx}. Inspired by these works in Refs.~\cite{Zhong:2021epq,Zhong:2022ecl}, we will restudy $D$-meson twist-2, 3 DAs $\phi_{2;D}(x,\mu)$, $\phi_{3;D}^p(x,\mu)$ and $\phi_{3;D}^\sigma(x,\mu)$ in this work.

The rest of the paper are organized as follows. In Sec.~\ref{sec:II}, we will present new sum rule formulae for the $\xi$-moments of $\phi_{2;D}(x,\mu)$, $\phi_{3;D}^p(x,\mu)$ and $\phi_{3;D}^\sigma(x,\mu)$, and briefly describe and improve the LCHO models of those DAs. In Sec.~\ref{sec:III}, we will analyze the behavior of those DAs based on the new values of $\xi$-moments in detail. Section~\ref{sec:IV} is reserved for a summery.

\section{Theoretical framework}\label{sec:II}

\subsection{New sum rule formulae for the $\xi$-moments of $D$-meson twist-2,3 DAs}
As discussion in Ref.~\cite{Zhong:2021epq}, the new sum rule formula for the $\xi$-moments is based on that the sum rule of zeroth moment can not be normalized in entire Borel parameter region. Therefore, the discussion of this paper begins with the sum rule formulae for the $\xi$-moments of $D$-meson twist-2 DA $\phi_{2;D}(x,\mu)$ obtained in Ref.~\cite{Zhang:2017rwz} and twist-3 DAs $\phi_{3;D}^p(x,\mu)$, $\phi_{3;D}^\sigma(x,\mu)$ obtained in Ref.~\cite{Zhong:2018exo}.

By giving up the priori setting for zeroth $\xi$-moment normalization, Eq. (28) in Ref.~\cite{Zhang:2017rwz} should be modified as
\begin{eqnarray} &&
\left<\xi^n\right>_{2;D}\left<\xi^0\right>_{2;D} \nonumber\\ &&
= \frac{M^2 e^{m_D^2/M^2}}{f_D^2} \left\{ \frac{1}{\pi} \frac{1}{M^2} \int^{s_D}_{m_c^2} ds e^{-s/M^2} {\rm Im} I_{\rm pert}(s) \right. \nonumber\\ &&
+ \hat{L}_M I_{\left<\bar{q}q\right>}(-q^2) + \hat{L}_M I_{\left<G^2\right>}(-q^2) + \hat{L}_M I_{\left<\bar{q}Gq\right>}(-q^2) \nonumber\\ &&
+ \left. \hat{L}_M I_{\left<\bar{q}q\right>^2}(-q^2) + \hat{L}_M I_{\left<G^3\right>}(-q^2) \right\},
\label{xinxi0}
\end{eqnarray}
for the $n$th $\xi$-moment $\left<\xi^n\right>_{2;D}$ of $\phi_{2;D}(x,\mu)$. Eq. (27) in Ref.~\cite{Zhong:2018exo} should be modified as
\begin{eqnarray} &&
\left<\xi^n_p\right>_{3;D}\left<\xi^0_p\right>_{3;D} \nonumber\\ &&
= \frac{M^2 e^{m_D^2/M^2}}{(\mu_D^p)^2 f_D^2} \left\{ \frac{1}{\pi} \frac{1}{M^2} \int^{s_D}_{m_c^2} ds e^{-s/M^2} {\rm Im} I^p_{\rm pert}(s) \right. \nonumber\\ &&
+ \hat{L}_M I^p_{\left<\bar{q}q\right>}(-q^2) + \hat{L}_M I^p_{\left<G^2\right>}(-q^2) + \hat{L}_M I^p_{\left<\bar{q}Gq\right>}(-q^2) \nonumber\\ &&
+ \left. \hat{L}_M I^p_{\left<\bar{q}q\right>^2}(-q^2) + \hat{L}_M I^p_{\left<G^3\right>}(-q^2) \right\},
\label{xinxi0p}
\end{eqnarray}
for the $n$th $\xi$-moment $\left<\xi^n_p\right>_{3;D}$ of $\phi_{3;D}^p(x,\mu)$. Eq. (28) in Ref.~\cite{Zhong:2018exo} should be modified as
\begin{eqnarray} &&
\left<\xi^n_\sigma\right>_{3;D}\left<\xi^0_p\right>_{3;D} \nonumber\\ &&
= \frac{3M^2 e^{m_D^2/M^2}}{(n+1)\mu_D^p \mu_D^\sigma f_D^2} \frac{m_D^2}{m_D^2-m_c^2} \nonumber\\ &&
\times \left\{ \frac{1}{\pi} \frac{1}{M^2} \int^{s_D}_{m_c^2} ds e^{-s/M^2} {\rm Im} I^\sigma_{\rm pert}(s) + \hat{L}_M I^\sigma_{\left<\bar{q}q\right>}(-q^2) \right. \nonumber\\ &&
+ \hat{L}_M I^\sigma_{\left<G^2\right>}(-q^2) + \hat{L}_M I^\sigma_{\left<\bar{q}Gq\right>}(-q^2) + \hat{L}_M I^\sigma_{\left<\bar{q}q\right>^2}(-q^2) \nonumber\\ &&
+ \left. \hat{L}_M I^\sigma_{\left<G^3\right>}(-q^2) \right\},
\label{xinxi0s}
\end{eqnarray}
for the $n$th $\xi$-moment $\left<\xi^n_\sigma\right>_{3;D}$ of $\phi_{3;D}^\sigma(x,\mu)$. In Eqs.~\eqref{xinxi0},~\eqref{xinxi0p} and~\eqref{xinxi0s}, $m_D$ is the $D$-meson mass, $m_c$ is the current charm-quark mass, $f_D$ is the decay constant of $D$-meson, $s_D$ is the continuum threshold, $\hat{L}_M$ indicates Borel transformation operator with the Borel parameter $M$. $\mu_D^p$ and $\mu_D^\sigma$ are the normalization constants of DAs $\phi_{3;D}^p(x,\mu)$ and $\phi_{3;D}^\sigma(x,\mu)$ respectively. Usually, $\mu_D^p = \mu_D^\sigma = \mu_D = m_D^2/m_c$ in literature by employing the equations of motion of on-shell quarks in the meson. However, as discussed in Refs.~\cite{Huang:2004tp,Huang:2005av}, the quarks inside the bound state are not exactly on-shell. Then $\mu_D^p$ and $\mu_D^\sigma$ are taken as undetermined parameters in this paper and will be determined via the sum rules of zeroth $\xi$-moments of DAs $\phi_{3;D}^p(x,\mu)$ and $\phi_{3;D}^\sigma(x,\mu)$ following the idea of Refs.~\cite{Zhong:2018exo,Huang:2004tp,Huang:2005av}. In addition, in sum rules \eqref{xinxi0}, \eqref{xinxi0p} and \eqref{xinxi0s}, the subscript ``pert'' stands for the terms coming from the contribution of perturbative part in operator product expansion, subscripts $\left<\bar{q}q\right>$, $\left<G^2\right>$, $\left<\bar{q}Gq\right>$, $\left<\bar{q}q\right>^2$ and $\left<G^3\right>$ stand for the terms proportional to double-quark condensate, double-gluon condensate, quark-gluon mixing condensate, four-quark condensate and triple-gluon condensate, respectively. For the expressions of those terms in Eqs.~\eqref{xinxi0},~\eqref{xinxi0p} and \eqref{xinxi0s}, one can refer to the appendixes in Refs.~\cite{Zhang:2017rwz,Zhong:2018exo}. By taking $n = 0$ in Eq.~\eqref{xinxi0} and \eqref{xinxi0p}, one can obtain the sum rules for the zeroth $\xi$-moments $\left<\xi^0\right>_{2;D}$ and $\left<\xi^0_p\right>_{3;D}$. As the functions of the Borel parameter, the zeroth $\xi$-moments $\left<\xi^0\right>_{2;D}$ in Eq.~\eqref{xinxi0} and $\left<\xi^0_p\right>_{3;D}$ in Eqs.~\eqref{xinxi0p} and \eqref{xinxi0s} obviously can not be normalized in entire $M^2$ region. Therefore, more reasonable and accurate sum rules should be
\begin{eqnarray}
\left<\xi^n\right>_{2;D} &=& \frac{\left<\xi^n\right>_{2;D}\left<\xi^0\right>_{2;D} |_{\rm From\ Eq.~\eqref{xinxi0}}}{\sqrt{\left<\xi^0\right>^2_{2;D}} |_{\rm By\ taking\ n=0\ in\ Eq.~\eqref{xinxi0}}}, \label{xin} \\
\left<\xi^n_p\right>_{3;D} &=& \frac{\left<\xi^n_p\right>_{3;D}\left<\xi^0_p\right>_{3;D} |_{\rm From\ Eq.~\eqref{xinxi0p}}}{\sqrt{\left<\xi^0_p\right>^2_{3;D}} |_{\rm By\ taking\ n=0\ in\ Eq.~\eqref{xinxi0p}}}, \label{xinp}
\end{eqnarray}
and
\begin{eqnarray}
\left<\xi^n_\sigma\right>_{3;D} = \frac{\left<\xi^n_\sigma\right>_{3;D}\left<\xi^0_p\right>_{3;D} |_{\rm From\ Eq.~\eqref{xinxi0s}}}{\sqrt{\left<\xi^0_p\right>^2_{3;D}} |_{\rm By\ taking\ n=0\ in\ Eq.~\eqref{xinxi0p}}}, \label{xins}
\end{eqnarray}
for $\left<\xi^n\right>_{2;D}$, $\left<\xi^n_p\right>_{3;D}$ and $\left<\xi^n_\sigma\right>_{3;D}$, respectively.

\subsection{LCHO models for $D$-meson twist-2, 3 DAs}\label{sec:IIB}

In Refs.~\cite{Zhang:2017rwz, Zhong:2018exo}, we have suggested LCHO models for $D$-meson twist-2, 3 DAs. In this subsection, we first propose a brief review for those models, then we will improve them by reconstructing whose longitudinal distribution functions.

The $D$-meson leading-twist DA $\phi_{2;D}(x,\mu)$ can be obtained by integrating out the transverse momentum $\textbf{k}_\perp$ component in its WF $\Psi_{2;D}(x,\textbf{k}_\perp)$, i.e.,
\begin{eqnarray}
\phi_{2;D}(x,\mu_0) = \frac{2\sqrt{6}}{f_D} \int_{\left|\textbf{k}_\perp\right|^2 \leq \mu_0^2} \frac{d^2\textbf{k}_\perp}{16\pi^3} \Psi_{2;D}(x,\textbf{k}_\perp).
\label{phiPsi2D}
\end{eqnarray}
Based on the BHL description~\cite{BHL}, the LCHO model for the $D$-meson leading-twist WF consists of the spin-space WF $\chi_{2;D}(x,\textbf{k}_\perp)$ and spatial WF $\psi_{2;D}^R(x,\textbf{k}_\perp)$, i.e., $\Psi_{2;D}(x,\textbf{k}_\perp) = \chi_{2;D}(x,\textbf{k}_\perp) \psi_{2;D}^R(x,\textbf{k}_\perp)$. The spin-space WF $\chi_{2;D}(x,\textbf{k}_\perp) = \widetilde{m}/\sqrt{\textbf{k}_\perp^2 + \widetilde{m}}$. In which, $\widetilde{m} = \hat{m}_c x + \hat{m}_q \bar{x}$ with the constituent charm-quark mass $\hat{m}_c$ and light-quark mass $\hat{m}_q$. In this paper, we take $\hat{m}_c = 1.5 {\rm GeV}$ and $\hat{m}_q = 0.25 {\rm GeV}$~\cite{Zhong:2022ecl}. As discussed in Ref.~\cite{Zhang:2017rwz}, we take $\chi_{2;D} \to 1$ approximately due to that $\hat{m}_c \gg \Lambda_{\rm QCD}$. Then, the $D$-meson leading-twist WF reads
\begin{eqnarray}
\Psi_{2;D}(x,\textbf{k}_\perp) &=& A_{2;D} \varphi_{2;D}(x) \nonumber\\
&\times& \exp \left[ -\frac{1}{\beta_{2;D}^2} \left( \frac{\textbf{k}_\perp^2 + \hat{m}_c^2}{\bar{x}} + \frac{\textbf{k}_\perp^2 + \hat{m}_q^2}{x} \right) \right], \nonumber\\
\label{psiR2D}
\end{eqnarray}
where $\bar{x} = 1-x$, $A_{2;D}$ is the normalization constant, $\beta_{2;D}$ is a harmonious parameter that dominates the WF's transverse distribution, $\varphi_{2;D}(x,\mu)$ dominates the WF's longitudinal distribution.

Substituting Eq.~\eqref{psiR2D} into \eqref{phiPsi2D}, the expression of $D$-meson leading-twist DA $\phi_{2;D}(x,\mu_0)$ can be obtained, i.e.,
\begin{eqnarray}
\phi_{2;D}(x,\mu) &=& \frac{\sqrt{6}A_{2;D}\beta_{2;D}^2}{\pi^2 f_D} x\bar{x} \varphi_{2;D}(x) \nonumber\\
&\times& \exp \left[ - \frac{\hat{m}_c^2x + \hat{m}_q^2\bar{x}}{8\beta_{2;D}^2 x\bar{x}} \right] \nonumber\\
&\times& \left\{ 1 - \exp \left[ - \frac{\mu^2}{8\beta_{2;D}^2 x\bar{x}} \right] \right\}.
\label{phi2D}
\end{eqnarray}
Following the way for constructing the $D$-meson leading-twist DA, the LCHO models for $D$-meson twist-3 DAs $\phi_{3;D}^p(x,\mu)$ and $\phi_{3;D}^\sigma(x,\mu)$ read,
\begin{eqnarray}
\phi_{3;D}^p(x,\mu) &=& \frac{\sqrt{6}A_{3;D}^p (\beta_{3;D}^p)^2}{\pi^2 f_D} x\bar{x} \varphi_{3;D}^p(x) \nonumber\\
&\times& \exp \left[ - \frac{\hat{m}_c^2x + \hat{m}_q^2\bar{x}}{8(\beta_{3;D}^p)^2 x\bar{x}} \right] \nonumber\\
&\times& \left\{ 1 - \exp \left[ - \frac{\mu^2}{8(\beta_{3;D}^p)^2 x\bar{x}} \right] \right\},
\label{phi3Dp}
\end{eqnarray}
and
\begin{eqnarray}
\phi_{3;D}^\sigma(x,\mu) &=& \frac{\sqrt{6}A_{3;D}^\sigma (\beta_{3;D}^\sigma)^2}{\pi^2 f_D} x\bar{x} \varphi_{3;D}^\sigma(x) \nonumber\\
&\times& \exp \left[ - \frac{\hat{m}_c^2x + \hat{m}_q^2\bar{x}}{8(\beta_{3;D}^\sigma)^2 x\bar{x}} \right] \nonumber\\
&\times& \left\{ 1 - \exp \left[ - \frac{\mu^2}{8(\beta_{3;D}^\sigma)^2 x\bar{x}} \right] \right\},
\label{phi3Ds}
\end{eqnarray}
respectively.

For the longitudinal distribution functions $\varphi_{2;D}(x)$, $\varphi_{3;D}^p(x)$ and $\varphi_{3;D}^\sigma(x)$, we used to take the first five terms of Gegenbauer expansions for the corresponding DAs in Refs.~\cite{Zhang:2017rwz, Zhong:2018exo}. As discussed in Ref.~\cite{Zhong:2021epq, Zhong:2022ecl}, higher order Gegenbauer polynomials will introduce spurious oscillations~\cite{Chang:2013pq}, while those corresponding coefficients obtained by directly solving the constraints of Gegenbauer moments or $\xi$-moments are not reliable. Then we improve these three longitudinal distribution functions as following,
\begin{eqnarray}
\varphi_{2;D}(x) &=& \left[ x(1-x) \right]^{\alpha_{2;D}} \left[ 1 + \hat{B}_1^{2;D} C_1^{3/2}(2x-1) \right], \nonumber\\ \label{varphi2D}\\
\varphi_{3;D}^p(x) &=& \left[ x(1-x) \right]^{\alpha_{3;D}^p} \left[ 1 + \hat{B}_{1,p}^{3;D} C_1^{1/2}(2x-1) \right], \nonumber\\ \label{varphi3Dp}\\
\varphi_{3;D}^\sigma(x) &=& \left[ x(1-x) \right]^{\alpha_{3;D}^\sigma} \left[ 1 + \hat{B}_{1,\sigma}^{3;D} C_1^{3/2}(2x-1) \right]. \nonumber\\ \label{varphi3Ds}
\end{eqnarray}

By considering the normalization conditions for $D$-meson twist-2, 3 DAs $\phi_{2;D}(x,\mu)$, $\phi_{3;D}^p(x,\mu)$ and $\phi_{3;D}^p(x,\mu)$, that is,
\begin{eqnarray}
\int^1_0 dx \phi_{2;D}(x,\mu) &=& \int^1_0 dx \phi_{3;D}^p(x,\mu) \nonumber\\
&=& \int^1_0 dx \phi_{3;D}^\sigma(x,\mu) = 1,
\label{DA_constraint1}
\end{eqnarray}
there are three undetermined parameters in the LCHO models for DAs $\phi_{2;D}(x,\mu)$, $\phi_{3;D}^p(x,\mu)$ and $\phi_{3;D}^p(x,\mu)$ respectively, and which will be taken as the fitting parameters to fit the first ten $\xi$-moments\footnote{In our previous work~\cite{Zhong:2022lmn}, based on the pionic leading-twist DA, we analyzed in detail the influence of different number of $\xi$-moments included in the fitting, and found that when the order of $\xi$-moments is not more than ten, the change of the number of $\xi$-moments has an obvious impact on the fitting results. When the order of $\xi$-moments is more than ten, the change of the number of $\xi$-moments has a very small impact on the fitting results. Therefore, we only use the first ten $\xi$-moments of $D$-meson DAs $\phi_{2;D}(x,\mu)$, $\phi_{3;D}^p(x,\mu)$ and $\phi_{3;D}^\sigma(x,\mu)$ for fitting in this work.} of corresponding DAs by adopting the least squares method in next section.

It should be noted that, $D$-meson twist-2, 3 DAs are the universal non-perturbative parameters in essence, and non-perturbative QCD should be used to study them in principle. However, due to the difficulty of non-perturbative QCD, those DAs are studied in this paper by combining the phenomenological model, that is, the LCHO model, and the non-perturbative QCD SRs in the framework of BFT. Otherwise, the improvement of the LCHO model of DAs $\phi_{2;D}(x,\mu)$, $\phi_{3;D}^p(x,\mu)$ and $\phi_{3;D}^p(x,\mu)$, that is, to reconstruct their longitudinal distribution functions, is only based on mathematical considerations. The rationality of this improvement can be judged by the goodness of fit.

\begin{table*}[htb]
\caption{Criteria for determining the Borel windows of the first ten $\xi$-moments of $D$-meson twist-2, 3 DAs. }
\begin{tabular}{ c  c  c  c  c  c }
\hline
~~ & ~~~~~~~Continue~~~~~~~ & ~~~~~~~Dimension-six~~~~~~~ & ~~~~~~~~~~~~~~~~~~~~ & ~~~~~~~Continue~~~~~~~ & ~~~~~~~Dimension-six~~~~~~~\\
~~ & ~Contribution ($\%$)~ & ~Contribution ($\%$)~ & ~~ & ~Contribution ($\%$)~ & ~Contribution ($\%$)~ \\
\hline
~$\left<\xi^1\right>_{2;D}$~ & ~$<15$~ & ~$<10$~ & ~$\left<\xi^2\right>_{2;D}$~ & ~$<20$~ & ~$<10$~ \\
~$\left<\xi^3\right>_{2;D}$~ & ~$-$~ & ~$<15$~ & ~$\left<\xi^4\right>_{2;D}$~ & ~$<20$~ & ~$<15$~ \\
~$\left<\xi^5\right>_{2;D}$~ & ~$-$~ & ~$<20$~ & ~$\left<\xi^6\right>_{2;D}$~ & ~$<20$~ & ~$<20$~ \\
~$\left<\xi^7\right>_{2;D}$~ & ~$-$~ & ~$<25$~ & ~$\left<\xi^8\right>_{2;D}$~ & ~$<20$~ & ~$<20$~ \\
~$\left<\xi^9\right>_{2;D}$~ & ~$-$~ & ~$<25$~ & ~$\left<\xi^{10}\right>_{2;D}$~ & ~$<20$~ & ~$<20$~ \\
\hline
~$\left<\xi^1_p\right>_{3;D}$~ & ~$-$~ & ~$<5$~ & ~$\left<\xi^2_p\right>_{3;D}$~ & ~$<15$~ & ~$<10$~ \\
~$\left<\xi^3_p\right>_{3;D}$~ & ~$-$~ & ~$<10$~ & ~$\left<\xi^4_p\right>_{3;D}$~ & ~$<15$~ & ~$<10$~ \\
~$\left<\xi^5_p\right>_{3;D}$~ & ~$-$~ & ~$<15$~ & ~$\left<\xi^6_p\right>_{3;D}$~ & ~$<15$~ & ~$<10$~ \\
~$\left<\xi^7_p\right>_{3;D}$~ & ~$-$~ & ~$<15$~ & ~$\left<\xi^8_p\right>_{3;D}$~ & ~$<15$~ & ~$<10$~ \\
~$\left<\xi^9_p\right>_{3;D}$~ & ~$-$~ & ~$<15$~ & ~$\left<\xi^{10}_p\right>_{3;D}$~ & ~$<15$~ & ~$<10$~ \\
\hline
~$\left<\xi^1_\sigma\right>_{3;D}$~ & ~$<20$~ & ~$-$~ & ~$\left<\xi^2_\sigma\right>_{3;D}$~ & ~$<45$~ & ~$-$~ \\
~$\left<\xi^3_\sigma\right>_{3;D}$~ & ~$<20$~ & ~$<5$~ & ~$\left<\xi^4_\sigma\right>_{3;D}$~ & ~$<45$~ & ~$-$~ \\
~$\left<\xi^5_\sigma\right>_{3;D}$~ & ~$<25$~ & ~$<5$~ & ~$\left<\xi^6_\sigma\right>_{3;D}$~ & ~$<45$~ & ~$-$~ \\
~$\left<\xi^7_\sigma\right>_{3;D}$~ & ~$<25$~ & ~$<5$~ & ~$\left<\xi^8_\sigma\right>_{3;D}$~ & ~$<45$~ & ~$-$~ \\
~$\left<\xi^9_\sigma\right>_{3;D}$~ & ~$<25$~ & ~$<5$~ & ~$\left<\xi^{10}_\sigma\right>_{3;D}$~ & ~$<45$~ & ~$-$~ \\
\hline
\end{tabular}
\label{tbwcri}
\end{table*}

\begin{figure*}[htb]
\centering
\includegraphics[width=0.32\textwidth]{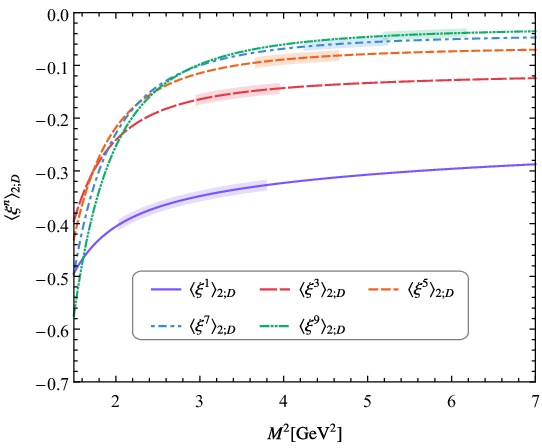}
\includegraphics[width=0.32\textwidth]{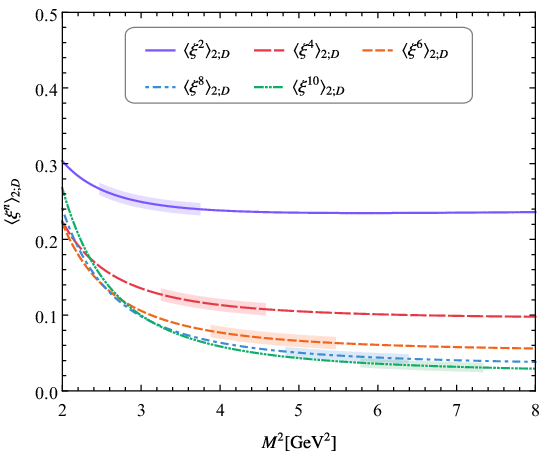}
\includegraphics[width=0.32\textwidth]{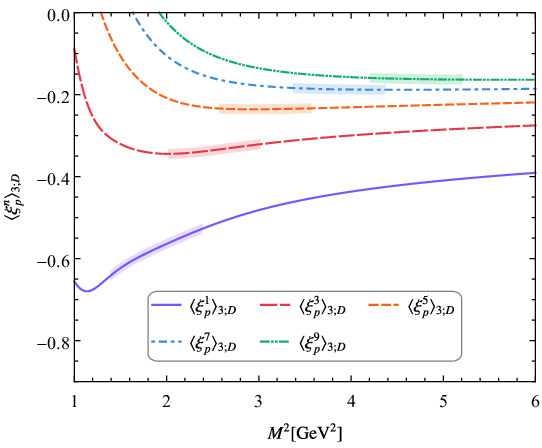}
\includegraphics[width=0.32\textwidth]{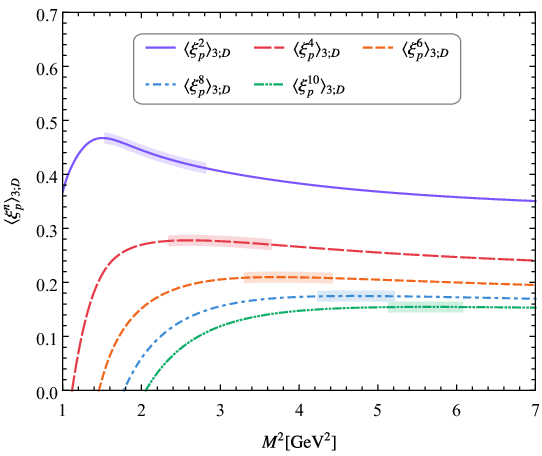}
\includegraphics[width=0.32\textwidth]{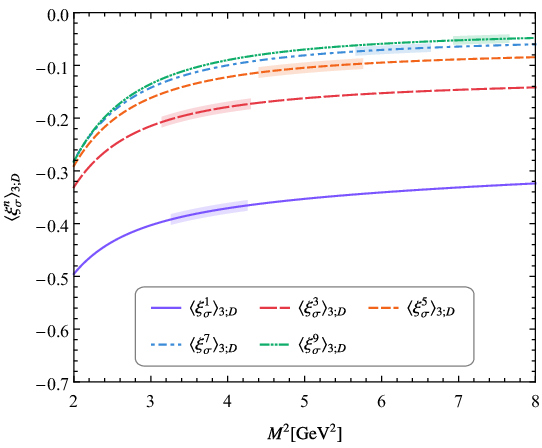}
\includegraphics[width=0.32\textwidth]{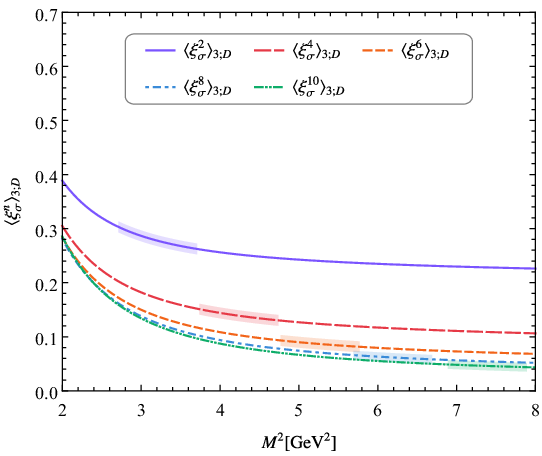}
\caption{The $D$-meson twist-2, 3 DA $\xi$-moments $\langle \xi^n\rangle _{2;D}$, $\langle \xi^n_p\rangle _{3;D}$ and $\langle \xi^n_\sigma\rangle _{3;D}$ with $(n=1,\cdots,10)$ versus the Borel parameter $M^2$. In order to clearly show the curves of different $\xi$-moments, only the central values of $\xi$-moments are given, which is obtained by taking the central values of each input parameters shown in Eqs.~\eqref{input1} and \eqref{input2}.}
\label{fxin}
\end{figure*}

\begin{table}[htb]
\caption{The first ten $\xi$-moments of the $D$-meson twist-2,~3 DAs $\phi_{2;D}(x,\mu)$, $\phi_{3;D}^p(x,\mu)$ and $\phi_{3;D}^\sigma(x,\mu)$ at scale $\mu = 2~{\rm GeV}$, respectively.}\label{table:xin_value}
\begin{tabular}{l l l l}
\hline\hline
$\left<\xi^1\right>_{2;D}$~~~~~~ & $-0.349^{+0.037}_{-0.037}$~~~~~~~~~ & $\left<\xi^2\right>_{2;D}$~~~~~~ & $0.251^{+0.014}_{-0.014}$ \\
$\left<\xi^3\right>_{2;D}$ & $-0.152^{+0.012}_{-0.012}$ & $\left<\xi^4\right>_{2;D}$ & $0.117^{+0.010}_{-0.010}$ \\
$\left<\xi^5\right>_{2;D}$ & $-0.0883^{+0.0072}_{-0.0072}$ & $\left<\xi^6\right>_{2;D}$ & $0.0715^{+0.0084}_{-0.0084}$ \\
$\left<\xi^7\right>_{2;D}$ & $-0.0606^{+0.0054}_{-0.0054}$ & $\left<\xi^8\right>_{2;D}$ & $0.0479^{+0.0052}_{-0.0052}$ \\
$\left<\xi^9\right>_{2;D}$ & $-0.0429^{+0.0034}_{-0.0034}$ & $\left<\xi^{10}\right>_{2;D}$ & $0.0348^{+0.0035}_{-0.0035}$ \\
\hline
$\left<\xi^1_p\right>_{3;D}$ & $-0.555^{+0.063}_{-0.061}$ & $\left<\xi^2_p\right>_{3;D}$ & $0.430^{+0.034}_{-0.036}$ \\
$\left<\xi^3_p\right>_{3;D}$ & $-0.325^{+0.026}_{-0.023}$ & $\left<\xi^4_p\right>_{3;D}$ & $0.272^{+0.017}_{-0.020}$ \\
$\left<\xi^5_p\right>_{3;D}$ & $-0.232^{+0.019}_{-0.017}$ & $\left<\xi^6_p\right>_{3;D}$ & $0.209^{+0.014}_{-0.017}$ \\
$\left<\xi^7_p\right>_{3;D}$ & $-0.185^{+0.019}_{-0.017}$ & $\left<\xi^8_p\right>_{3;D}$ & $0.175^{+0.016}_{-0.017}$ \\
$\left<\xi^9_p\right>_{3;D}$ & $-0.163^{+0.017}_{-0.016}$ & $\left<\xi^{10}_p\right>_{3;D}$ & $0.157^{+0.015}_{-0.016}$ \\
\hline
$\left<\xi^1_\sigma\right>_{3;D}$ & $-0.376^{+0.021}_{-0.021}$ & $\left<\xi^2_\sigma\right>_{3;D}$ & $0.280^{+0.023}_{-0.023}$ \\
$\left<\xi^3_\sigma\right>_{3;D}$ & $-0.188^{+0.020}_{-0.019}$ & $\left<\xi^4_\sigma\right>_{3;D}$ & $0.141^{+0.012}_{-0.013}$ \\
$\left<\xi^5_\sigma\right>_{3;D}$ & $-0.1078^{+0.0104}_{-0.0103}$ & $\left<\xi^6_\sigma\right>_{3;D}$ & $0.0890^{+0.0077}_{-0.0079}$ \\
$\left<\xi^7_\sigma\right>_{3;D}$ & $-0.0735^{+0.0055}_{-0.0054}$ & $\left<\xi^8_\sigma\right>_{3;D}$ & $0.0635^{+0.0054}_{-0.0055}$ \\
$\left<\xi^9_\sigma\right>_{3;D}$ & $-0.0550^{+0.0037}_{-0.0036}$ & $\left<\xi^{10}_\sigma\right>_{3;D}$ & $0.0489^{+0.0041}_{-0.0042}$ \\
\hline\hline
\end{tabular}
\label{txin}
\end{table}

\section{numerical analysis}\label{sec:III}

\subsection{Inputs}

To do the numerical calculation for the $\xi$-moments of $D$-meson twist-2, 3 DAs, we take the scale $\mu = M$ as usual, and take $\Lambda_{\rm QCD}^{(n_f)} \simeq 324, 286, 207~{\rm MeV}$ for the number of quark flavors $n_f = 3, 4, 5$, respectively~\cite{Zhong:2021epq,Zhong:2022ecl}. For other inputs, we take~\cite{Workman:2022ynf}
\begin{eqnarray}
m_{D^-} &=& 1869.66 \pm 0.05 {\rm MeV}, \nonumber\\
f_D &=& 203.7 \pm 4.7 \pm 0.6 {\rm MeV}, \nonumber\\
\bar{m}_c(\bar{m}_c) &=& 1.27 \pm 0.02 {\rm GeV}, \nonumber\\
m_d(2{\rm GeV}) &=& 4.67^{+0.48}_{-0.17} {\rm MeV}, \label{input1}
\end{eqnarray}
and~\cite{Zhong:2021epq, Zhong:2014jla, Colangelo:2000dp}
\begin{eqnarray}
\left<\bar{q}q\right>(2{\rm GeV}) &=& \left( -2.417^{+0.227}_{-0.114} \right) \times 10^{-2} {\rm GeV}^3, \nonumber\\
\left<g_s\bar{q}\sigma TGq\right>(2{\rm GeV}) &=& \left( -1.934^{+0.188}_{-0.103} \right) \times 10^{-2} {\rm GeV}^5, \nonumber\\
\left<g_s\bar{q}q\right>^2(2{\rm GeV}) &=& \left( 2.082^{+0.734}_{-0.697} \right) \times 10^{-3} {\rm GeV}^6, \nonumber\\
\left<\alpha_sG^2\right> &=& 0.038 \pm 0.011 {\rm GeV}^4, \nonumber\\
\left<g_s^3fG^3\right> &=& 0.045 {\rm GeV}^6. \label{input2}
\end{eqnarray}
The renormalization group equations of those inputs are~\cite{Zhong:2021epq}
\begin{eqnarray}
m_d (\mu) &=& m_d (\mu_0) \left[ \frac{\alpha_s(\mu_0)}{\alpha_s(\mu)} \right]^{-4/\beta_0}, \nonumber\\
\bar{m}_c (\mu) &=& \bar{m}_c (\bar{m}_c) \left[ \frac{\alpha_s(\bar{m}_c)}{\alpha_s(\mu)} \right]^{-4/\beta_0}, \nonumber\\
\langle \bar{q}q\rangle (\mu) &=& \langle \bar{q}q\rangle (\mu_0) \left[ \frac{\alpha_s(\mu_0)}{\alpha_s(\mu)} \right]^{4/\beta_0}, \nonumber\\
\langle g_s\bar{q}\sigma TGq\rangle (\mu) &=& \langle g_s\bar{q}\sigma TGq\rangle (\mu_0) \left[ \frac{\alpha_s(\mu_0)}{\alpha_s(\mu)} \right]^{-2/(3\beta_0)}, \nonumber\\
\langle g_s\bar{q}q\rangle^2 (\mu) &=& \langle g_s\bar{q}q\rangle^2 (\mu_0) \left[ \frac{\alpha_s(\mu_0)}{\alpha_s(\mu)} \right]^{4/\beta_0}, \nonumber\\
\langle \alpha_s G^2\rangle (\mu) &=& \langle \alpha_s G^2\rangle (\mu_0), \nonumber\\
\langle g_s^3fG^3\rangle (\mu) &=& \langle g_s^3fG^3\rangle (\mu_0),
\label{RGE}
\end{eqnarray}
with $\beta_0 = (33-2n_f)/3$. For the continuum threshold, we used to take $s_D \simeq 6.5 {\rm GeV}^2$ in Ref.~\cite{Zhang:2017rwz, Zhong:2018exo}. This value comes from the square of the mass of $D$-meson's first exciting state, i.e., $D^0(2550)$, as suggested by Refs.~\cite{BaBar:2010zpy,Li:2012gr}. In Refs.~\cite{Zhong:2021epq,Zhong:2022ecl}, we take the continuum threshold parameters $s_\pi$ and $s_K$ by requiring that there are reasonable Borel windows to normalize the zeroth $\xi$-moments of the pion and kaon leading-twist DAs. In this paper, we follow the suggestion in Refs.~\cite{Zhong:2021epq,Zhong:2022ecl}, and get $s_D \simeq 6.0 {\rm GeV}^2$.

\subsection{$\xi$-moments and behaviors of $D$-meson twist-2, 3 DAs}

Then we can calculate the values of the $\xi$-moments of $D$-meson twist-2, 3 DAs with the sum rules \eqref{xin}, \eqref{xinp} and \eqref{xins}. First, one need to determine the appropriate Borel windows for those $\xi$-moments by following usual criteria, such as the contributions of continuum state and dimension-six condensate are as small as possible, and the values of those $\xi$-moments are stable in corresponding Borel windows. Table~\ref{tbwcri} exhibits the limits to the continuum state's contributions and the dimension-six condensate's contributions for the first ten $\xi$-moments of $D$-meson twist-2, 3 DAs. In which, the symbol ``$-$'' indicates that corresponding continuum state's contribution is smaller than $10\%$ or dimension-six condensate's contribution is much smaller than $5\%$ in a wide Borel parameter region. This is reasonable because both continuum state's contribution and dimension-six condensate's contribution are depressed by the sum rules of zeroth $\xi$-moments in the denominator of the new sum rule formulae~\eqref{xin}, \eqref{xinp} and \eqref{xins}. By comparing with the criteria listed in Table 1 and Table 4 in Ref.~\cite{Zhong:2018exo}, the criteria listed in Table~\ref{tbwcri} are much stricter, which reflects one of the advantages of the new sum rule formulae~\eqref{xin}, \eqref{xinp} and \eqref{xins}, that is, they reduce the system uncertainty of the sum rule itself. Then, for those $\xi$-moments, only the upper or lower limits of the corresponding Borel windows is clearly determined. In order to get complete Borel windows, we directly take their lengths as $1~{\rm GeV}^2$. Figure~\ref{fxin} shows the $D$-meson twist-2, 3 DA $\xi$-moments $\langle \xi^n\rangle _{2;D}$, $\langle \xi^n_p\rangle _{3;D}$ and $\langle \xi^n_\sigma\rangle _{3;D}$ with $(n=1,\cdots,10)$ versus the Borel parameter $M^2$. In this figure, the uncertainties caused by the errors of input parameters is not drawn to clearly show the curves of different $\xi$-moments. Meanwhile, the Borel windows are also shown with the shaded bands. By taking all error sources, such as $D$-meson mass and decay constant, $u$- and $c$-quark masses, as well as vacuum condensates, etc., shown in Eqs.~\eqref{input1} and \eqref{input2}, into consideration, and adding the uncertainties in quadrature, the values of the first ten $\xi$-moments of $D$-meson twist-2, 3 DAs are shown in Table~\ref{txin}. Here, we give the first two Gegenbauer moments of $D$-meson twist-2, 3 DAs for reference, that is,
\begin{eqnarray}
a_1^{2;D} &=& -0.582^{+0.062}_{-0.062}, \quad\quad a_2^{2;D} = 0.148^{+0.042}_{-0.042}, \nonumber\\
a_{1,p}^{3;D} &=& -1.665^{+0.188}_{-0.182}, \quad\quad a_{2,p}^{3;D} = 0.726^{+0.252}_{-0.273}, \nonumber\\
a_{1,\sigma}^{3;D} &=& -0.626^{+0.036}_{-0.035}, \quad\quad a_{2,\sigma}^{3;D} = 0.232^{+0.067}_{-0.068},
\label{an}
\end{eqnarray}
at scale $\mu = 2{\rm GeV}$.

In the above work, in order to calculate the $\xi$-moments of $D$-meson twist-3 DAs $\phi_{3;D}^p(x,\mu)$ and $\phi_{3;D}^\sigma(x,\mu)$, one should calculate the normalization constants $\mu_D^p$ and $\mu_D^\sigma$ first. Under the assumption that the sum rules of zeroth $\xi$-moments $\left<\xi^0_p\right>_{3;D}$ and $\left<\xi^0_\sigma\right>_{3;D}$ can be normalized in appropriate Borel windows, the sum rules of $\mu_D^p$ and $\mu_D^\sigma$ can be obtained by taking $n=0$ in Eqs.~\eqref{xinp} and \eqref{xins} and substituting $\left<\xi^0_p\right>_{3;D} = \left<\xi^0_\sigma\right>_{3;D} = 1$ into these two sum rules. We require the continuum state's contributions are less than $30\%$ and dimension-six condensate's contributions are not more than $5\%$ and $0.5\%$ to determine the Borel windows for $\mu_D^p$ and $\mu_D^\sigma$, respectively. By adding the uncertainties derived from all error sources in quadrature, we have,
\begin{eqnarray}
\mu_D^p = 2.717^{+0.087}_{-0.087}, \quad\quad \mu_D^\sigma = 2.231^{+0.073}_{-0.068},
\label{muD}
\end{eqnarray}
at scale $\mu = 2{\rm GeV}$. Compared with the values in Ref.~\cite{Zhong:2018exo}, $\mu_D^p$ in \eqref{muD} increases by about $7.2\%$, and $\mu_D^\sigma$ decreases by about $12.0\%$. The former is caused by the update of input parameters, while the latter is also caused by the new sum rule formula, i.e., Eq.~\eqref{xins}, in addition to the update of input parameters.

\begin{table*}[htb]
\caption{Goodness of fit and the Values of the LCHO model parameters for $D$-meson twist-2 DA $\phi_{2;D}(x,\mu)$ and twist-3 DAs $\phi_{3;D}^p(x,\mu)$ and $\phi_{3;D}^\sigma(x,\mu)$ at scale $\mu = 2~{\rm GeV}$.}
\begin{tabular}{l l l l l l}
\hline\hline
~$A_{2;D}~({\rm GeV}^{-1})$~~~~~~~~~~ & ~$\alpha_{2;D}$~~~~~~~~~~ & ~$B_1^{2;D}$~~~~~~~~~~ & ~$\beta_{2;D}~({\rm GeV})$~~~~~~~~~~ & ~$\chi^2_{\rm min}$~~~~~~~~~~ & ~$P_{\chi^2_{\rm min}}$ \\
~$34.4712$~ & ~$-0.861$~ & ~$0.107$~ & ~$0.535$~ & ~$0.873219$~ & ~$0.996623$~ \\
\hline
~$A_{3;D}^p~({\rm GeV}^{-1})$~~~~~~~~~~ & ~$\alpha_{3;D}^p$~~~~~~~~~~ & ~$B_{1,p}^{3;D}$~~~~~~~~~~ & ~$\beta_{3;D}^p~({\rm GeV})$~~~~~~~~~~ & ~$\chi^2_{\rm min}$~~~~~~~~~~ & ~$P_{\chi^2_{\rm min}}$ \\
~$0.536764$~ & ~$-1.360$~ & ~$-0.922$~ & ~$1.135$~ & ~$2.39892$~ & ~$0.934514$~ \\
\hline
~$A_{3;D}^\sigma~({\rm GeV}^{-1})$~~~~~~~~~~ & ~$\alpha_{3;D}^\sigma$~~~~~~~~~~ & ~$B_{1,\sigma}^{3;D}$~~~~~~~~~~ & ~$\beta_{3;D}^\sigma~({\rm GeV})$~~~~~~~~~~ & ~$\chi^2_{\rm min}$~~~~~~~~~~ & ~$P_{\chi^2_{\rm min}}$ \\
~$28.9986$~ & ~$-1.403$~ & ~$0.228$~ & ~$0.484$~ & ~$0.594628$~ & ~$0.999021$~ \\
\hline\hline
\end{tabular}
\label{t:model_parameter}
\end{table*}

\begin{figure*}[htb]
\centering
\includegraphics[width=0.32\textwidth]{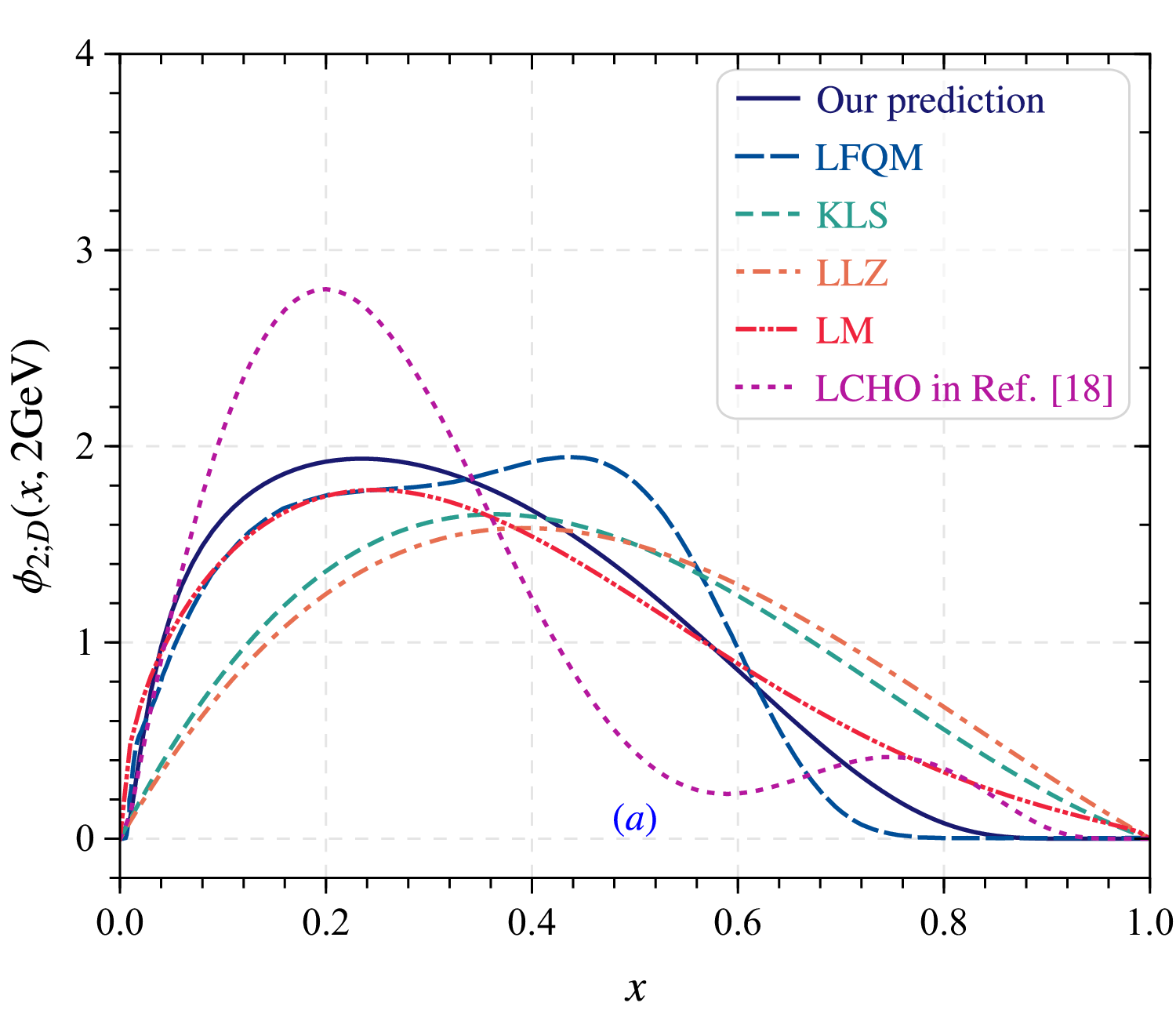}
\includegraphics[width=0.32\textwidth]{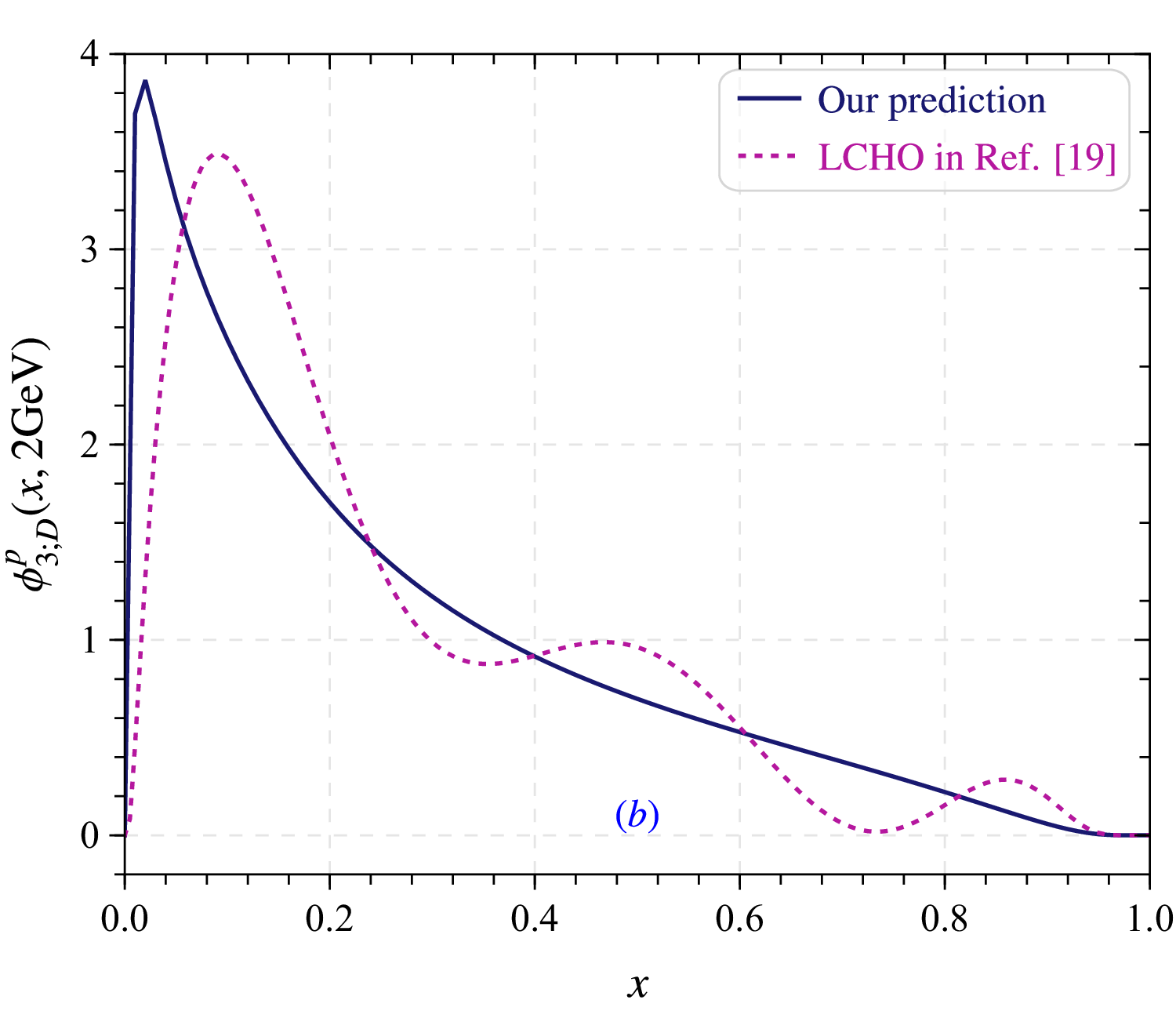}
\includegraphics[width=0.32\textwidth]{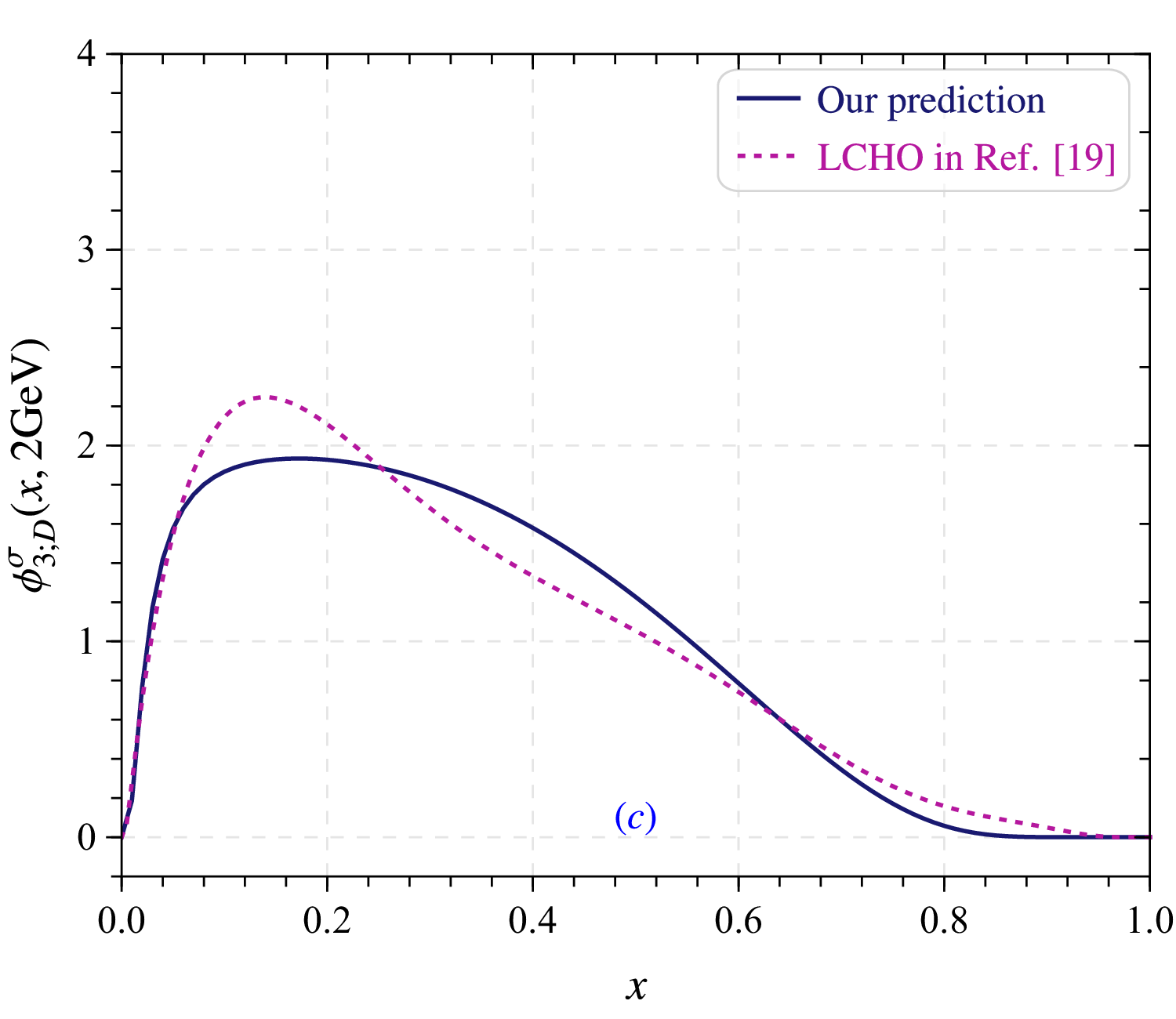}
\caption{Curves of the $D$-meson twist-2 DA $\phi_{2;D}(x,\mu)$ and twist-3 DAs $\phi_{3;D}^p(x,\mu)$ and $\phi_{3;D}^\sigma(x,\mu)$ at scale $\mu = 2~{\rm GeV}$. The models in literature such as KLS model~\cite{Kurimoto:2002sb}, LLZ model~\cite{Li:2008ts}, LM model~\cite{Li:1999kna}, the form with LFQM~\cite{Dhiman:2019ddr}, and our previous research results~\cite{Zhang:2017rwz, Zhong:2018exo} based on the LCHO model are also shown for comparison.}
\label{f:DAs}
\end{figure*}

Then we can determine the model parameters of our LCHO models for $D$-meson twist-2 DA $\phi_{2;D}(x,\mu)$ and twist-3 DAs $\phi_{3;D}^p(x,\mu)$ and $\phi_{3;D}^\sigma(x,\mu)$ by using the $\xi$-moments exhibited in Table~\ref{txin} with the least squares method following the way suggested in Refs.~\cite{Zhong:2021epq,Zhong:2022ecl}. Take the $D$-meson leading-twist DA $\phi_{2;D}(x,\mu)$ as an example, we first take the fitting parameters $\theta$ as the undetermined LCHO model parameters $\alpha_{2;D}$, $B_1^{2;D}$ and $\beta_{2;D}$, i.e., $\theta = (\alpha_{2;D}, B_1^{2;D}, \beta_{2;D})$, as discussed in Sec.~\ref{sec:IIB}. By minimizing the likelihood function
\begin{align}
\chi^2(\theta) = \sum^{10}_{i=1} \frac{(y_i - \mu(i,\theta))^2}{\sigma_i^2},
\label{eq:lf}
\end{align}
the optimal values of the fitting parameters $\theta$ we are looking for can be obtained. In Eq.~\eqref{eq:lf}, $i$ is taken to be the order of the $\xi$-moments of $\phi_{2;D}(x,\mu)$; the central values of $\xi$-moments $\langle\xi^n\rangle_{2;D} (n = 1,\cdots,10)$ with their errors exhibited in Table~\ref{table:xin_value} are regarded as the independent measurements $y_i$ and the corresponding variance $\sigma_i$. One can intuitively judge the goodness of fit through the magnitude of probability $P_{\chi^2_{\rm min}} = \int^\infty_{\chi^2_{\rm min}} f(y;n_d) dy$ with the probability density function of $\chi^2(\theta)$, i.e., $f(y; n_d) = \frac{1}{\Gamma(n_d/2) 2^{n_d/2}} y^{n_d/2 - 1} e^{-y/2}$, where $n_d$ indicates the number of degrees of freedom. The obtained optimal values of the model parameters $\alpha_{2;D}$, $B_1^{2;D}$ and $\beta_{2;D}$ at scale $\mu = 2~{\rm GeV}$ and the corresponding goodness of fit are shown in Table~\ref{t:model_parameter}. Following the same procedure, the LCHO model parameters for $D$-meson twist-3 DAs $\phi^p_{3;D}(x,\mu)$ and $\phi^\sigma_{3;D}(x,\mu)$ at scale $\mu = 2~{\rm GeV}$ and the corresponding goodness of fits can be obtained and are shown in Table~\ref{t:model_parameter} too. Then the corresponding behaviors of DAs $\phi_{2;D}(x,\mu)$, $\phi^p_{3;D}(x,\mu)$ and $\phi^\sigma_{3;D}(x,\mu)$ are determined. In order to intuitively show the behaviors of these three DAs, we plot and exhibit their curves in Fig.~\ref{f:DAs}. As a comparison, the models in literature for $D$-meson leading-twist DA $\phi_{2;D}(x,\mu)$ such as KLS model~\cite{Kurimoto:2002sb}, LLZ model~\cite{Li:2008ts}, LM model~\cite{Li:1999kna}, the form with LFQM~\cite{Dhiman:2019ddr}, and our previous research results~\cite{Zhang:2017rwz, Zhong:2018exo} for $\phi_{2;D}(x,\mu)$, $\phi^p_{3;D}(x,\mu)$ and $\phi^\sigma_{3;D}(x,\mu)$ based on the LCHO model are also shown in Fig.~\ref{f:DAs}. From Fig.~\ref{f:DAs}, one can find that our present prediction for $\phi_{2;D}(x,\mu)$ is closes to LM model. Compared with the KLS model and LLZ model, our $\phi_{2;D}(x,\mu)$ is narrower, and supports a large momentum distribution of valence quark in $x \sim [0.05, 0.5]$. Compared with our previous work in Refs.~\cite{Zhang:2017rwz, Zhong:2018exo}, our new predictions for $\phi_{2;D}(x,\mu)$, $\phi^p_{3;D}(x,\mu)$ and $\phi^\sigma_{3;D}(x,\mu)$ in this paper is smoother, and effectively eliminating the spurious oscillations introduced by the high-order Gegenbauer moments in old LCHO model.

\section{summary}\label{sec:IV}

In this paper, we restudied the $D$-meson leading-twist DA $\phi_{2;D}(x,\mu)$, twist-3 DAs $\phi_{3;D}^p(x,\mu)$ and $\phi_{3;D}^\sigma(x,\mu)$ with QCD SRs in the framework of BFT by adopting a new scheme suggested in our previous work~\cite{Zhong:2021epq}. The new sum rule formula for the $\xi$-moments $\langle\xi^n\rangle_{2;D}$, $\langle\xi^n_p\rangle_{3;D}$ and $\langle\xi^n_\sigma\rangle_{3;D}$, i.e., Eqs.~\eqref{xin},~\eqref{xinp} and~\eqref{xins}, were proposed and used to calculate whose values. Those values have been exhibited in Table~\ref{table:xin_value}. The LCHO models for DAs $\phi_{2;D}(x,\mu)$, $\phi_{3;D}^p(x,\mu)$ and $\phi_{3;D}^\sigma(x,\mu)$ were improved. By fitting the values of $\xi$-moments $\langle\xi^n\rangle_{2;D}$, $\langle\xi^n_p\rangle_{3;D}$ and $\langle\xi^n_\sigma\rangle_{3;D}$ shown in Table~\ref{table:xin_value} with the least squares method, the model parameters were determined and shown in Table~\ref{t:model_parameter}. Then the predicted curves for $D$-meson leading-twist DA $\phi_{2;D}(x,\mu)$, twist-3 DAs $\phi_{3;D}^p(x,\mu)$ and $\phi_{3;D}^\sigma(x,\mu)$ are shown in Fig.~\ref{f:DAs}.

The criteria adopted to determine the Borel windows for $\xi$-moments of $D$-meson leading-twist DA $\phi_{2;D}(x,\mu)$, twist-3 DAs $\phi_{3;D}^p(x,\mu)$ and $\phi_{3;D}^\sigma(x,\mu)$ exhibited in Table~\ref{tbwcri} imply that the new sum rule formula~\eqref{xin},~\eqref{xinp} and~\eqref{xins} can reduce the system uncertainties and propose more accurate predictions for $\xi$-moments $\langle\xi^n\rangle_{2;D}$, $\langle\xi^n_p\rangle_{3;D}$ and $\langle\xi^n_\sigma\rangle_{3;D}$, respectively. The goodness of fits for $\phi_{2;D}(x,\mu)$, $\phi_{3;D}^p(x,\mu)$ and $\phi_{3;D}^\sigma(x,\mu)$ are $P_{\chi^2_{\rm min}} = 0.996623$, $0.934514$ and $0.999021$, respectively, which indicate our improved LCHO models shown in Sec.~\ref{sec:IIB} with the model parameters in Table~\ref{t:model_parameter} can well prescribe the behaviors of those three DAs. The predicted DAs' curves shown in Fig.~\ref{f:DAs} indicate the improved LCHO models in this work can eliminate the spurious oscillations introduced by the high-order Gegenbauer moments in old LCHO models obtained in Refs.~\cite{Zhang:2017rwz, Zhong:2018exo}. Otherwise, in order to simply investigate the influence of the new $D$ meson twist-2, 3 DAs in this work on the relevant physical quantities, the TFFs $f_{+,0}^{B\to D}(q^2)$ and $\mathcal{R}(D)$ are calculated. For the relevant formulae, one can refer to Ref.~\cite{Zhong:2018exo}. We find that the new DAs can bring about $10\%$ and $6\%$ changes to $f_{+,0}^{B\to D}(0)$ and $\mathcal{R}(D)$ respectively. In order to obtain a more accurate TFFs and $\mathcal{R}(D)$, it is necessary to consider the next-to-leading order corrections for the contributions of $D$ meson twist-3 DAs, which will be our next step.

\section{Acknowledgments}
This work was supported in part by the National Natural Science Foundation of China under Grant No.12265009 and No.12265010, the Project of Guizhou Provincial Department of Science and Technology under Grant No.ZK[2021]024 and No.ZK[2023]142, the Project of Guizhou Provincial Department of Education under Grant No.KY[2021]030, and by the Chongqing Graduate Research and Innovation Foundation under Grant No. ydstd1912.

\end{document}